\documentclass[a4paper,12pt]{article}
\usepackage{epsfig}
\usepackage{amssymb}
\usepackage{amsfonts}
\usepackage{amsmath}
\usepackage{euscript}
\usepackage{verbatim}
\usepackage{latexsym}
\usepackage{graphicx}

\newskip\humongous \humongous=0pt plus 1000pt minus 1000pt

\newif\ifdtup

\catcode`\@=11

\@addtoreset{equation}{section}

\def\@normalsize{\@setsize\normalsize{15pt}\xiipt\@xiipt
\abovedisplayskip 14pt plus3pt minus3pt%
\belowdisplayskip \abovedisplayskip
\abovedisplayshortskip \z@ plus3pt%
\belowdisplayshortskip 7pt plus3.5pt minus0pt}

\def\small{\@setsize\small{13.6pt}\xipt\@xipt
\abovedisplayskip 13pt plus3pt minus3pt%
\belowdisplayskip \abovedisplayskip
\abovedisplayshortskip \z@ plus3pt%
\belowdisplayshortskip 7pt plus3.5pt minus0pt
\def\@listi{\parsep 4.5pt plus 2pt minus 1pt
     \itemsep \parsep
     \topsep 9pt plus 3pt minus 3pt}}

\relax

\catcode`@=12

\catcode`\@=11

\def\section{\@startsection{section}{1}{\z@}{3.5ex plus 1ex minus
   .2ex}{2.3ex plus .2ex}{\large\bf}}

\def\SymBoxes#1#2#3#4{\newdimen\un@t \un@t#3%
\raisebox{#1}{\rule{#2\un@t}{#4}\hskip-#2\un@t% lower horizontal
\@tempdimb\un@t \advance\@tempdimb by-#4\@tempcntb#2\relax%
\@whilenum{\@tempcntb>0}\do{%                         % #2 vertical lines
\rule{#4}{\un@t}\hskip\@tempdimb \advance\@tempcntb by\m@ne}%
\hskip-#2\un@t \rule[\un@t]{#2\un@t}{#4}%
\rule[\un@t]{#4}{#4}\hskip-#4%             % upper horizontal line
\rule{#4}{\un@t}}\hskip-#4}                % rightest vertical line
\begin{document}
%\begin{letter}{~}

%%%%%%Define some new commands and  macros
\newcommand{\beq}{\begin{equation}}
\newcommand{\eeq}{\end{equation}}
\newcommand{\bea}{\begin{eqnarray}}
\newcommand{\eea}{\end{eqnarray}}
\newcommand{\beas}{\begin{eqnarray*}}
\newcommand{\eeas}{\end{eqnarray*}}
\newcommand{\defi}{\stackrel{\rm def}{=}}
\newcommand{\non}{\nonumber}
\newcommand{\bquo}{\begin{quote}}
\newcommand{\enqu}{\end{quote}}
%%%%%%%%%%%%%%%%
\renewcommand{\(}{\begin{equation}}
\renewcommand{\)}{\end{equation}}
%%%%%%%%%%%%%%%%%%%%%%%%%%%%%%%%%% definitions
\def \eqn#1#2{\begin{equation}#2\label{#1}\end{equation}}
\def\IZ{{\mathbb Z}}
\def\IR{{\mathbb R}}
\def\IC{{\mathbb C}}
\def\IQ{{\mathbb Q}}
\def\de{\partial}
\def\Tr{ \hbox{\rm Tr}}
\def\H{ \hbox{\rm H}}
\def\HE{ \hbox{$\rm H^{even}$}}
\def\HO{ \hbox{$\rm H^{odd}$}}
\def\K{ \hbox{\rm K}}
\def\Im{ \hbox{\rm Im}}
\def\Ker{ \hbox{\rm Ker}}
\def\const{\hbox {\rm const.}}
\def\o{\over}
\def\im{\hbox{\rm Im}}
\def\re{\hbox{\rm Re}}
\def\bra{\langle}\def\ket{\rangle}
\def\Arg{\hbox {\rm Arg}}
\def\Re{\hbox {\rm Re}}
\def\Im{\hbox {\rm Im}}
\def\exo{\hbox {\rm exp}}
\def\diag{\hbox{\rm diag}}
\def\longvert{{\rule[-2mm]{0.1mm}{7mm}}\,}
\def\a{\alpha}
\def\dag{{}^{\dagger}}
\def\tq{{\widetilde q}}
\def\p{{}^{\prime}}
\def\W{W}
\def\N{{\cal N}}
\def\hsp{,\hspace{.7cm}}

\def\br{\nonumber\\}
\def\IZ{{\mathbb Z}}
\def\IR{{\mathbb R}}
\def\IC{{\mathbb C}}
\def\IQ{{\mathbb Q}}
\def\IP{{\mathbb P}}
\def \eqn#1#2{\begin{equation}#2\label{#1}\end{equation}}

\newcommand{\C}{\ensuremath{\mathbb C}}
\newcommand{\Z}{\ensuremath{\mathbb Z}}
\newcommand{\R}{\ensuremath{\mathbb R}}
\newcommand{\rp}{\ensuremath{\mathbb {RP}}}
\newcommand{\cp}{\ensuremath{\mathbb {CP}}}
\newcommand{\vac}{\ensuremath{|0\rangle}}
\newcommand{\vact}{\ensuremath{|00\rangle}                    }
\newcommand{\oc}{\ensuremath{\overline{c}}}
\begin{titlepage}
\begin{flushright}
%CHEP XXXXX
%ULB-TH/09-10\\
%hep-th/yymmnnn\\
\end{flushright}
\bigskip
\def\thefootnote{\fnsymbol{footnote}}

\begin{center}
{\Large
{\bf A Kaluza-Klein Subttractor
}
}
\end{center}

\bigskip
\begin{center}
{\large Sanjib JANA$^a$\footnote{{\texttt{sanjib@cts.iisc.ernet.in}}} and Chethan KRISHNAN$^a$\footnote{{\texttt{chethan@cts.iisc.ernet.in}}}
%Avinash RAJU$^a$\footnote{{\texttt{avinashraju777@gmail.com}}} \\
%\vspace{0.1in}
%and Somyadip THAKUR$^a$\footnote{{\texttt{smydp3thkr@gmail.com}}} 
}
\vspace{0.1in}

\end{center}

\renewcommand{\thefootnote}{\arabic{footnote}}

\begin{center}
%\vspace{0.2cm}
$^a$ {Center for High Energy Physics\\
Indian Institute of Science, Bangalore, India}\\

\end{center}

\noindent
\begin{center} {\bf Abstract} \end{center}
We generalize the results of arXiv:1212.1875 and arXiv:1212.6919 on attraction basins and their boundaries to the case of a specific class of rotating black holes, namely the ergo-free branch of extremal black holes in Kaluza-Klein theory. We find that exact solutions that span the attraction basin can be found even in the rotating case by appealing to certain symmetries of the equations of motion. They are characterized by two asymptotic parameters that generalize those of the non-rotating case, and the boundaries of the basin are spinning versions of the (generalized) subttractor geometry. We also give examples to illustrate that the shape of the attraction basin can drastically change depending on the theory.

\vspace{1.6 cm}
\vfill

\end{titlepage}

\setcounter{footnote}{0}

%%%%%%%%%%%%%%%%%%%%%%%%%%%%%%%%%%%%%%%%%%%%%%%%%%%%%%%%%%%%%%%%%%%%%%%%%%%%%%%%%%%%%%%%%%%%%%
%%%%%%%%%%%%%%%%%%%%%%%%%%%%%%%%%%%%%%%%%%%%%%%%%%%%%%%%%%%%%%%%%%%%%%%%%%%%%%%%%%%%%%%%%%%%%%
\section{Introduction}
\label{intro}

Attractor mechanism \cite{FKS} is the statement that the moduli scalars in an extremal black hole geometry approach fixed values at the horizon even though their radial profiles can be different. In some recent work \cite{A1, A2} we demonstrated that the boundary of the basin of attraction of a class of static attractor black holes in dilatonic supergravity is the so-called (generalized) subtracted geometry \cite{CL1, CL2, CG} in the extremal limit. We called these solutions subttractors\footnote{Subtracted geometries initially arose in the context of the Kerr/CFT correspondence \cite{KerrCFT} and as a geometrical way to manifest the hidden conformal symmetry of black holes \cite{HCS}. Subtracted geometry has been constructed and further discussed in \cite{CL1, CL2, CG, A1, A2, General1}.}.

In this paper, we wish to generalize this statement to a class of examples where the extremal black hole is rotating, and determine the boundary of the attraction basin in this more general case as well. This is non-trivial because adding rotation breaks the spherical symmetry of the problem down to an axial symmetry. % and makes the problem substantially more complicated. 

The theory we will work with is the simplest Kaluza-Klein theory, namely Einstein-Maxwell-dilaton theory obtained by reduction of the 5D Einstein action \cite{Pope}. In this paper we will study (rotating) extremal dyonic black holes of this theory with electric and magnetic charges. In section 2, we will show how in the static case, the equations of motion of this theory can be fully integrated and the extremal solutions are characterized by two parameters (integration constants) which we call $d_1$ and $d_2$. The basin of attraction of asymptotically flat dyons is described by a certain domain in the $d_1, d_2$ space (which we describe). This domain of $d_1, d_2$ that we find here is quite different from the attraction basin found in the theories in \cite{A1, A2}. To emphasize that the domains of attraction and the structure of the solutions can vary drastically, in an appendix\footnote{The theory considered in this appendix is a dilatonic toy model, independent of the KK theory we discuss in the rest of the paper.} we also discuss this domain for yet another fully solvable system (case III in \cite{Goldstein}) and describe its basin structure.

In section 3, we will consider the same Kaluza-Klein theory, but now we will look at what we are really after: rotating (extremal) solutions. Once we turn on rotation, the system is substantially more complicated. The black hole solutions of this theory were written down in \cite{Rasheed, Matos, LarsenKK}. There are two kinds of extremal limits that these black holes have, the so called ergo-branch and the ergo-free branch \cite{Kevin}. The branch that allows a natural generalization from the static case is the ergo-free branch, and that is the one we will study. As it turns out, the black hole solutions of \cite{Rasheed, Matos, LarsenKK} are not general enough to discuss the attraction basin that generalizes the static basin of section 2. Fortunately,  despite the complications of the rotating case, using some symmetries of the equations of motion we are able to generate precisely such solutions. These new solutions are again characterized by two parameters which we call $d_1, d_2$ in analogy with the static case. 

We conclude in section 4 by discussing the basin of attraction of these new solutions and presenting the full attractor flow structure for various values of $\theta$, the polar angle in the geometry. The boundaries of attraction are characteried again by certain limits of the parameters $d_1$ and $d_2$. The discussion of these boundaries is somewhat distracting, so we have relegated it to an appendix. In these limits, the solutions degenerate and turn into a generalization of the (rotating) subttractor geometry. The warp factor of the geometry goes as $\sim r^3$ for these solutions, which should be contrasted to $\sim r^4$ for flat space and $\sim r$ in the original subtracted geometry of \cite{CL1, CL2, CG}.  

{\bf Note added:} After this work was substantially completed, two papers appeared on the arXiv which deal with related ideas. \cite{CveticStudent} constructed\footnote{Generalizing \cite{JdBoer}.} general black hole flow solutions and \cite{Torino} deals with extremal black holes in dilatonic supergravity. These works rely on the Harrison transfromation machinery\footnote{See \cite{Amitabh, Klemm, General1} for previous discussions on Harrison transformations in the subtracted geometry context.} to generate solutions. These solutions should enable an exhaustive study of the attraction basin structure of general (extremal) black holes \cite{General} in these theories, going far beyond our approach here, which is based on symmetry arguments and other parlor tricks.

\section{The Static Attractor: Kaluza-Klein Dyon}

The reduction of 5D Einstein action on a circle gives \cite{Pope}
\bea
S=\int d^4x \sqrt{-g}\Big(R-2 (\partial \phi)^2- e^{2 \sqrt{3} \phi} F_{\mu\nu}F^{\mu\nu} \Big). \label{action}
\eea
In this section we will look for general static extremal solutions of this theory. The static ``attractor ansatz" that we will use is
\bea
ds^2=-a(r)^2 dt^2+\frac{dr^2}{a(r)^2}+b(r)^2 d \Omega^2, \hspace{1in} \\
F=Q_1\ \sin \theta d\theta \wedge\ d\phi+ \frac{ e^{-2 \sqrt{3} \phi}}{b^2} Q_2 \ dt \wedge dr,\ \ \ \ \phi \equiv \phi(r). \label{gauge-scalar}\hspace{ 0.1in}
\eea
The resulting equations of motion are
\bea
(a^2\ b^2)''-2 =0 \label{att1} \\
\frac{b''}{b}+{\phi'}^2=0 \\
(a^2 b^2\phi')'-\frac{\partial_{\phi} V_{\rm eff}(\phi)}{2 b^2}=0
\eea
and the energy constraint from Einstein equations is
\bea 
a^2{b'}^2+\frac{{a^2}'{b^2}'}{2}+\frac{V_{\rm eff}(\phi)}{b^2}-a^2b^2 {\phi'}^2-1=0. \label{minansatze2}
\eea
The effective potential $V_{\rm eff}$ is
\bea\label{mineff}
V_{\rm eff}(\phi)=e^{2 \sqrt{3} \phi}Q_1^2+e^{-2 \sqrt{3} \phi}Q_2^2. \label{effpot}
\eea
Here, we obtained the equations of motion and the effective potential by using an electric plus magnetic (ie., dyonic) ansatz in a theory with one gauge field. It is straightforward to see that one can also get these exact same equations, by (say) starting with a theory where instead we have two gauge fields with suitable couplings to one scalar, with only magnetic charges turned on. 

These equations can in fact be solved exactly \cite{Goldstein, GibMaed, Pope2}. This has been done in the appendix of \cite{Goldstein} after imposing the condition that the system has a double-zero horizon (ie., it is extremal). But they make the further assumption that the asymptotic value of $a(r)$ is unity, which is how asymptotic flatness is imposed. We find it useful for our purposes to leave this freedom arbitrary. This has a few virtues. One is that this essentially gives us the most general solution of the system which is extremal: in other words all the near-horizon data has been fixed, but none of the boundary data is. The second virtue is that in many cases, we need this degree of freedom to gain a full understanding of the attraction basin. Finally, it is evident from the equations of motion above that there is a rescaling symmetry to the system under 
\bea
r \rightarrow r/\mu, \ \ a \rightarrow a /\mu, 
\eea
In effect, fixing the value of $a(r)$ at $r \rightarrow \infty $ fixes this scale.  Not fixing this scaling is tantamount to leaving it as an integration constant in the solutions. This is what we will do, because a generalization of this scaling symmetry exists also in the rotating case as well and we will make use of it in generating our new solutions there. This is crucial because unlike in the static case, solving the system frontally is an industry in itself in the rotating case: the equations are complicated partical differential equations and the system has less symmetry.

It is straightforward to construct the general solution by adapting the results in Appendix B.3.2 of \cite{Goldstein}. The result can be expressed as
\bea
e^{2\sqrt{3} \phi}=\frac{Q_2}{Q_1}\ \left[\frac{2 (d_2^2-d_1)r^2-2d_2 r +1}{2d_1 r^2-2d_2 r +1}\right]^{3/2}, \hspace{1.2in}\\
a^2(r)=\frac{1}{2 Q_1 Q_2}\frac{r^2}{\sqrt{(2 (d_2^2-d_1^2)r^2-2d_2 r +1)(2d_1 r^2-2d_2 r +1)}}, \ \ b(r)=r/a(r).
\eea
Here $d_2$ and $d_1$ are the integration constants, and the asymptotic values of $a(r)$ and $\phi(r)$ (lets call them $a_0$ and $\phi_\infty$) can be immediately read off from the above expressions in terms of them. The results of \cite{Goldstein} were obtained for the special case, $a_0=1$. 

Now, the attraction basin structure can be easily studied using the approach spelled out in \cite{A2}. The idea is that we want the solutions to be regular in the region $0<r<\infty$ and this will happen only if the quadratic polynomial expressions that show up in the solution above have no zeros in that range. The zeros (in $r$) of the polynomials can be determined in terms of $d_1$ and $d_2$ and this imposes inequalities on the ranges of values that $d_2$ and $d_1$ can take. We present the analyses of these inequalities in an appendix, the final result is that the attraction basin is the acute region 
\bea
d_1>0, \ \ d_2<0, \ \ {\rm and} \ \ d_2^2 >d_1, \label{statbasin}
\eea
in the $d_1- d_2$ plane that is bounded by the curves
\bea
d_1=0, \ \ d_1=d_2^2, \ \ {\rm with} \ \ d_2 <0.\label{statbound}
\eea    
This is the hatched region in figure 1. Note that the attraction basin is no longer a quadrant as it was in the cases considered in \cite{A2}. 
\begin{figure}
\begin{center}
\includegraphics[height=0.30\textheight]{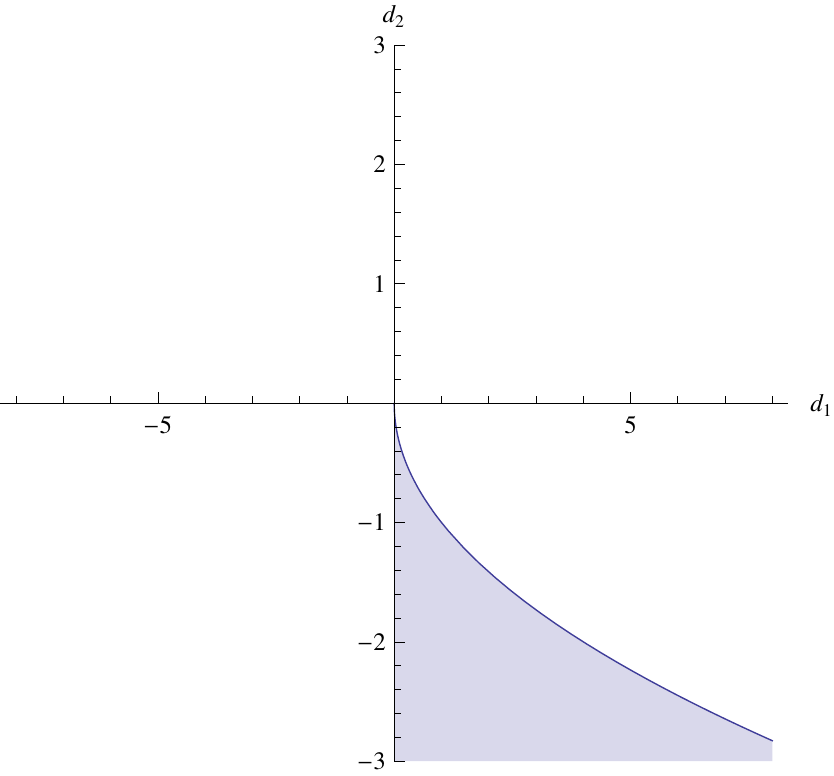}
\caption{The attraction basin of the Kaluza-Klein black hole. The basin structure in terms of $d_1$ and $d_2$ is identical, whether the black hole is spinnning or not. The hatched region is bounded by the curves $d_2^2=d_1$ and $d_1=0$.}
\end{center}
\end{figure}
To emphasize that the curves defining the attraction boundaries can take other forms, we consider another exactly solvable dilatonic gravity theory in an appendix and describe its attraction basin in the static case.

If we slice through the attraction basin\footnote{We will use the phrase ``attraction basin" to refer to the hatched region in the $d_1 - d_2$ parameter space.} along a $d_2={\rm constant}$ line, we will see the various asymptotically flat hairy solutions that arise as we go from one boundary of the attraction basin to the other. The resulting plot is identical to the plots with $\beta=0$ or $\theta=\pi/2$ in figure 2, so we will not repeat them here. If we perturb a black hole with parameters $d_1=d_1^0$ and $d_2=d_2^0$ near the horizon by spherically symmetric perturbations and evolve it radially, we will find another solution in the same family with a different value of $d_1$ and $d_2$. In fact this new solution and old solution will lie precisely on the line
\bea
d_2= {\rm negative \  constant} =d_2^0 \label{flowline2}
\eea
in the attraction basin. The perturbation theory is straightforward (even if tedious) to do by a simple adaptation of the results in Appendix D of \cite{A2} (see also very similar computations in \cite{SCK1, SCK2, SCK3} for some more detail). We will not emphasize it here, because we are merely interested in {\em some} illustrative slicing of the attraction basin\footnote{The slicing emerging from the perturbation theory is perhaps the most canonical slicing of the attractor basin, however.}. The only salient point is that because of the specific form of the couplings in our system, the leading non-zero term in in the scalar perturbation appears not at ${\cal O}(r)$, but ${\cal O}(r^2)$, see eg. \cite{A1}. It turns out that this perturbation (lets call it $\phi_2$) is the only degree of freedom for radial fluctuations, and it therefore parametrizes a line along the basin:
\bea
d_1=d_1^0-\frac{\phi_{2}}{\sqrt{3}}, \ \ d_2=d_2^0. 
\eea
(Trivially) eliminating $\phi_2$ from system is how we obtained (\ref{flowline2}).

At the boundaries of attraction, the solutions degenerate. At the $d_1=0$ broundary it becomes
\bea
e^{2\sqrt{3} \phi}=\frac{Q_2}{Q_1}\ \left[\frac{2 d_2^2 r^2-2d_2 r +1}{1-2d_2 r}\right]^{3/2}, \hspace{1in}\\
a^2(r)=\frac{1}{2 Q_1 Q_2}\frac{r^2}{\sqrt{(2 d_2^2 r^2-2d_2 r +1)(1-2d_2 r )}}, \ \ b(r)=r/a(r).
\eea
and at the $d_2^2=d_1$ end, the solution is
\bea
e^{2\sqrt{3} \phi}=\frac{Q_2}{Q_1}\ \left[\frac{1-2d_2 r}{2d_2^2 r^2-2d_2 r +1}\right]^{3/2}, \hspace{1in}\\
a^2(r)=\frac{1}{2 Q_1 Q_2}\frac{r^2}{\sqrt{(1-2d_2 r)(2d_2^2 r^2-2d_2 r +1)}}, \ \ b(r)=r/a(r).
\eea
The warp factor $\Delta$ defined by $\sqrt{\Delta}\equiv r^2/a^2$ determines the asymptotic behavior of the geometry, and it goes as $\sim r^3$ in both cases. So the solutions in this limit stop being asymptotically flat, and are examples of subtracted geometries \cite{CL1, CG, JdBoer, A1, A2}. Note that the two solutions above are essentially identical except for a sign flip (and shift) in the scalar.

\section{Adding Rotation: the Ergo-Free Branch}

Adding rotation makes the black hole significantly more complicated. Rotating Kaluza-Klein black holes were first constructed in \cite{Rasheed, Matos, LarsenKK}. In the non-extremal case, on top of a mass parameter $M_k$ and the charges $Q_1$ and $Q_2$, now we have a rotation parameter $a_k$ in the metric \cite{Kevin}. Since our interest is in the attractor mechanism, we will only be interested in extremal solutions and will not write down the general non-extremal metric.

Starting with the non-extremal case, there are two kinds of extremal limits one can take. One involves setting $a_k=M_k$. This is called the ergo branch because the solution has an ergosphere. The other extremal limit is taken by sending $a_k, M_k \rightarrow 0$ while holding $a_k/M_k \equiv \beta$ fixed. This limit is called the ergo-free branch and this is what we will be interested in\footnote{In the non-rotating case, the extremal limit in these coordinates corresponds to sending $M_k \rightarrow 0$. So the ergo-free branch is the natural generalization of the static extremal limit. Understanding the attraction basins of all possible extremal solutions is likely to require full control on all the hairy solutions. This should be possible to do in light of the recent results in \cite{CveticStudent, Torino}, but we will not attempt it here.}. Regularity of the solution dictates that $|\beta|<1$ \cite{Rasheed}. Note that the sign of $\beta$ merely captures the direction of spin.

Our starting point will be the hairless extremal solutions discussed in \cite{Rasheed, LarsenKK, Kevin}. We will write the metric using the electric and magnetic charges ($Q, P$) in the normalization of \cite{Rasheed}. In the ergo-free extremal limit, $a_k$ and $M_k$ do not show up because they have gone to zero, but their ratio $\beta$ doees. The metric takes the form:
\bea
ds^2-\frac{X}{\sqrt{f_1 f_2}}(dt - \omega d\varphi)^2+\frac{\sqrt{f_1 \ f_2}}{X}dr^2+\sqrt{f_1 f_2} d\Omega^2 \label{metric}
\eea
with
\bea
f_1=r^2+\frac{P^{2/3}\sqrt{P^{2/3}+Q^{2/3}}}{2 \sqrt{\pi}}r+\frac{P^{4/3}Q^{2/3}}{8 \pi} (1-\beta \cos \theta),\label{mi} \\
f_2=r^2+\frac{Q^{2/3}\sqrt{P^{2/3}+Q^{2/3}}}{2 \sqrt{\pi}}r+\frac{P^{2/3}Q^{4/3}}{8 \pi} (1+\beta \cos \theta), \\
X=r^2, \ \ \omega=\frac{P Q}{8 \pi}\ \frac{\beta \sin^2 \theta}{r}. \label{mf}\hspace{1in}
\eea
The lack of spherical symmetry entrers the metric through the non-trivial angular dependence on the time fibration.

In the static case we could use an ansatz for the field strength which automatically solved the gauge field equations without using the details of the specific solution (cf. the attractor ansatz from the previous section). But now, we have to work with gauge fields directly and the non-zero components take the following form:
\bea
A_t=-\frac{Q}{4 \sqrt{\pi} \  f_2}\left[r+\frac{P^{2/3}\sqrt{P^{2/3}+Q^{2/3}}}{4 \sqrt{\pi}} (1+\beta \cos \theta)\right], \hspace{0.5in}\label{ai} \\
A_\varphi=-\frac{P}{4 \sqrt{\pi}}\left[\cos \theta+ \frac{\beta \sin^2 \theta}{2\ f_2}\Big(\frac{Q^{2/3}\sqrt{P^{2/3}+Q^{2/3}}}{2 \sqrt{\pi}}r+\frac{P^{2/3}Q^{4/3}}{4 \pi}\Big) \right]. \label{af}
\eea
Together with the scalar
\bea
\exp(2 \sqrt{3} \phi)=\Big(\frac{f_2}{f_1}\Big)^{\frac{3}{2}}, \label{s}
\eea
this completes the description of the black hole. 

Note that the solution above has no free parameters other than the charges of the hole (counting also $\beta$). What we would like is a generalization of this solution which has two free parameters (integration constants) generalizing what we saw in the static case, so that we can conveniently describe the attraction basin. One of these parameters corresponds to the freedom to scale the asymptotics of the metric and the other one is the freedom to shift the asymptotic value of the scalar. The latter is simple to implement \cite{Kevin}. We see from the action (\ref{action}) that the theory has a symmetry under 
\bea
\phi \rightarrow \phi+ \lambda, \ \ A_\mu \rightarrow e^{-\sqrt{3} \lambda} A_{\mu}.
\eea
This transformation however changes the charges of the black hole, but we can rescale the charge parameters appearing in the solution while doing the above shift/scaling to obtain a one-parameter family of solutions with identical electric magnetic charges \cite{Kevin}:
\bea
\phi \rightarrow \phi+ \lambda, \ \ A_\mu \rightarrow e^{-\sqrt{3} \lambda} A_{\mu}, \ \ Q \rightarrow e^{-\sqrt{3} \lambda} Q, \ \ P \rightarrow e^{\sqrt{3} \lambda} P. \label{lambda}
\eea
This is a (one parameter) solution generating transformation that we can use. 

But where does the second parameter come from? The hint here is that the static ansatz had a scaling symmetry in $a$ which could be removed by rescaling $r$. We will look for a generalization of this to the rotating case here. But since we do not have an ansatz when rotation is present\footnote{Naively adding arbitrary $\theta$ dependence to the ansatz functions results in PDEs. Moreover, choosing an ansatz for the gauge field will be non-trivial.  However, even though we have not explored this in detail, it seems possible that by choosing the $\theta$ dependence of the ansatz carefully and using inspired analogies from the exact solution \cite{Rasheed, Matos, LarsenKK}, one might get a tractable set of EoMs. The solution we find here is enough for our purposes since it is applicable in the extremal limit, but the  static ``attractor" ansatz is applicable even away from extremality.}, we cannot immediately search for this symmetry in the equations of motion. But an inspection of the metric (\ref{metric}) and the gauge field
\bea
A=A_t \ dt+ A_\varphi \ d\varphi
\eea
reveals that they (and trivially, the scalar) are invariant under 
\bea
r \rightarrow r/\mu,\ t \rightarrow t \ \mu,\ X \rightarrow  X/\mu^2,\ \omega \rightarrow \omega \ \mu,\  A_t \rightarrow A_t /\mu.
\eea
This means that if we have a solution, replacing $r$ by $r/\mu$ will result in a new solution if we compensate for it by an {\em opposite} scaling of $\omega, X$ and $A_t$.

Doing the above two transformations results in the new metric functions:
\bea
f_{1}=r^2/\mu^2+\frac{P_0^{2/3}\sqrt{P_0^{2/3}+Q_0^{2/3}}}{2 \sqrt{\pi}}r/\mu+\frac{P_0^{4/3}Q_0^{2/3}}{8 \pi} (1-\beta \cos \theta),\label{tempi} \\
f_{2}=r^2/\mu^2+\frac{Q_0^{2/3}\sqrt{P_0^{2/3}+Q_0^{2/3}}}{2 \sqrt{\pi}}r/\mu+\frac{P_0^{2/3}Q_0^{4/3}}{8 \pi} (1+\beta \cos \theta), \\
X=r^2, \ \ \omega=\frac{P_0 Q_0}{8 \pi}\ \frac{\beta \sin^2 \theta}{r}. \label{tempf}\hspace{1in}
\eea
The gauge field components (in terms of the {\em new} metric functions $f_1$ and $f_2$) take the form:
\bea
A_t=-\mu \frac{Q_0 e^{-\sqrt{3}\lambda}}{4 \sqrt{\pi} \  f_2}\left[r/\mu+\frac{P_0^{2/3}\sqrt{P_0^{2/3}+Q_0^{2/3}}}{4 \sqrt{\pi}} (1+\beta \cos \theta)\right], \hspace{0.5in}\label{tempai} \\
A_\varphi=-\frac{P_0 e^{-\sqrt{3}\lambda}}{4 \sqrt{\pi}}\left[\cos \theta+ \frac{\beta \sin^2 \theta}{2\ f_2}\Big(\frac{Q_0^{2/3}\sqrt{P_0^{2/3}+Q_0^{2/3}}}{2 \sqrt{\pi}}r/\mu+\frac{P_0^{2/3}Q_0^{4/3}}{4 \pi}\Big) \right]. \label{tempaf}
\eea
Finally, the scalar (again in terms of the new $f_1$ and $f_2$) is:
\bea
\exp(2 \sqrt{3} \phi)=\exp(2 \sqrt{3} \lambda)\Big(\frac{f_2}{f_1}\Big)^{\frac{3}{2}}. \label{news}
\eea 
It is important to note that in the above expressions, $P_0$ and $Q_0$ are the scaled charges:
\bea
Q_0 = e^{-\sqrt{3} \lambda} Q, \ \ P_0 = e^{\sqrt{3} \lambda} P.
\eea
These are the two parameter solutions we seek, and they are pretty ugly and complicated in terms of $\beta, P, Q, \mu$ and $\lambda$.

But since we have two free parameters at this stage, we can compare their static limit with the static solutions of section 2 and write them in a cleaner notation. It turns out that this notation also lends itself to a better description of the attraction basin. Comparing the expressions after setting $\beta=0$ in the formulas above, results in the relations
\bea
\lambda = \frac{1}{2 \sqrt{3}} \log\left[\frac{Q_2}{Q_1} \Big(\frac{d_2^2 - d_1}{d_1}\Big)^{3/2}\right], \ 
\mu = \frac{1}{2 \sqrt{Q_1 Q_2}} \left[\frac{1}{d_1 (d_2^2- d_1)}\right]^{1/4}, \\
\nonumber \\
P = -4 \sqrt{\pi} Q_1,  \ \ Q = -4 \sqrt{\pi} Q_2. \hspace{1.5in}
\eea
So apart from a minor difference in the normalization, the charges $P, Q$ of the black hole solutions of \cite{Rasheed, Kevin} are the same as in our notation, $Q_1, Q_2$. In terms of the $(Q_1, Q_2, d_1, d_2)$ notation, then the full rotating solutions take the much more tractable form
\bea
X(r) = r^2, \ \ \omega = \frac{2\ Q_1\ Q_2} {r}\beta \sin^2 \theta, \hspace{0.8in} \label{final1} \\ 
f_1 = 2\ Q_1\ Q_2\ \sqrt{\frac{d_2^2-d_1}{d_1}}
(2\ d_1\ r^2  - 2\ d_2\ r +1 - \beta \ \cos \theta), \hspace{0.05in} \\
f_2 = 2\ Q_1\ Q_2\ \sqrt{\frac{d_1}{d_2^2-d_1}} \left(2 (d_2^2-d_1)r^2-2d_2 r +1+\beta \cos \theta\right)  \label{final2}
\eea
with the gauge fields
\bea
A_t= \frac{Q_1}{2\ f_2}\sqrt{\frac{d_1}{(d_2^2-d_1)^3}}\Big[2 d_1 r-d_2(1+\beta \cos \theta)\Big], \hspace{0.3in}\\
A_\varphi=Q_1\left[\cos \theta-\frac{2 Q_1 Q_2}{f_2} \sqrt{\frac{d_1}{d_2^2-d_1}}(d_2 r -1)\beta \sin^2 \theta\right],
\eea
and the scalar
\bea
\exp(2 \sqrt{3} \phi)=\frac{Q_2}{Q_1}\Big(\frac{d_2^2 - d_1}{d_1}\Big)^{3/2}\Big(\frac{f_2}{f_1}\Big)^{\frac{3}{2}}.
\eea

We have checked explicitly that these comprise a solution of the full\footnote{As opposed to a solution of the attractor ansatz of section 2, which is only valid in the static limit.} set of equations of motion arising from the action (\ref{action}). This is the form of the solution that we will use to investigae the attraction basin in the next section.

\section{The Basin}

The solutions presented in the previosu section are regular only for specific ranges of the parameters $d_1$ and $d_2$. These ranges can be detremined by systematic application of the requirement of regularity, and is done in an appendix. The end result that one finds is what we will call the basin of attraction. It turns out that the basin for the rotating black hole for any value of $\theta$ is bounded by the same curves as in the static case (\ref{statbasin}-\ref{statbound}). We plot the solutions for a few values of  $\beta$ and $\theta$ in figure 2. 
\begin{figure}
\begin{center}
\includegraphics[height=0.40\textheight]{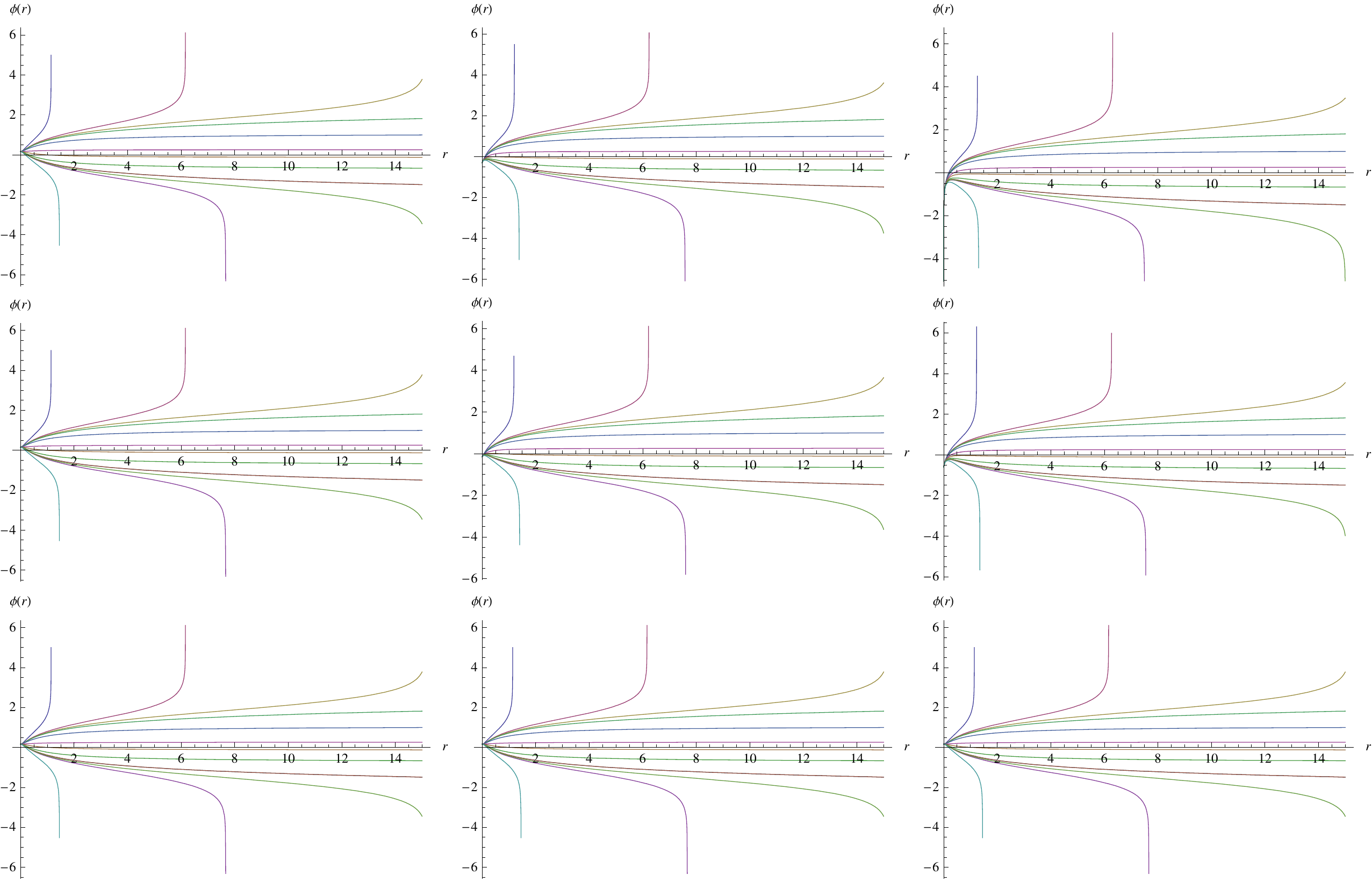}
\caption{Curves in each row is a fixed $\beta$($=0, 0.5, 1$, left-right), and each column is a fixed $\theta$($=0, \pi/4, \pi/2$, top-down). The last non-divergent line is the attraction boundary.}
\end{center}
\end{figure}
As expected, the plots with $\theta=\pi/2$ or $\beta=0$ give identical curves - these correspond to the attraction basin of the static attractor. 

It is worthwhile making a couple of comments about these solutions. %Firstly,  since it is a spinning black hole, the  solution should have an up-down symmetry about the equator. This is why we have only plotted the polar angle range between $0$ and $\pi/2$. 
The attraction basin structure is determined by the regularity of the functions $f_1$ and $f_2$ (see Appendix A). It is clear from the form of these functions that when 
\bea
\beta \cos \theta \rightarrow -\beta \cos \theta, \label{sym1}
\eea
this is tantamount to the switch
\bea
d_1 \rightarrow d_2^2-d_1, \ \ d_2 \rightarrow d_2. \label{sym2}
\eea
But under such a replacement, the attraction basin (\ref{statbasin}-\ref{statbound}) retains its structure. This means that we only have to determine the attraction basin for positive values of $\beta \cos \theta$.
%, and the scalar essentially changes only by a sign. The warp factor $f_1 f_2$ does not change. These are all indications of the fact that the underlying system is a rotating black hole and changing the sign of rotation $\beta$ (or equivalently, measuring the polar angle from the South pole which flips the sign of $\cos \theta$) should be a symmetry of the system. We will because that results in merely changing the direction of rotation (which is idnetical to taking the coordinates of the solution from the South pole asopposed to the North pole). But if one looks at the explicit functions in our solution, naively there is no symmetry because of the way $d_2$ and $d_1$ enter the solution asymmetrically. This puzzle is resolved, when one observes that the flip in the rotation (or equivalently in the hemisphere) 

At the boundaries of attraction:
\bea
f_1 f_2= 4 Q_1^2 Q_2^2 (-2d_2 \ r +1 -\beta \cos \theta)(2 d_2^2 \ r^2-2d_2 \ r+1+\beta \cos \theta), \\ 
\exp (2 \sqrt{3} \phi)=\frac{Q_2}{Q_1}\Big(\frac{2 d_2^2 \ r^2-2d_2 \ r+1+\beta \cos \theta}{-2d_2 \ r +1 -\beta \cos \theta}\Big)^{\frac{3}{2}}, \hspace{0.7in}
\eea
when $d_1=0$, and
\bea
f_1 f_2= 4 Q_1^2 Q_2^2 (2 d_2^2 \ r^2-2d_2 \ r+1- \beta \cos \theta)(-2d_2 \ r +1 +\beta \cos \theta), \\ 
\exp (2 \sqrt{3} \phi)=\frac{Q_2}{Q_1}\Big(\frac{-2d_2 \ r +1 +\beta \cos \theta}{2 d_2^2 \ r^2-2d_2 \ r+1- \beta \cos \theta}\Big)^{\frac{3}{2}}, \hspace{0.7in}
\eea
when $d_2^2=d_1$. From these it follows that the metric and the scalar both have well-defined limits at the boundaries. Asymptotically, the warp factor goes $\sim r^3$ as in the non-rotating case, so this is a rotating version of the (generalized) subttractor geometry found in \cite{A1, A2}. It is worth noting here that the solutions here again exhibit a symmetry under the flipping of the two boundaries, essentially upro a flip in the sign of the scalar and $\beta \cos \theta \rightarrow -\beta \cos \theta$. This is related to the discussion in the last paragraph.
 
We haven't explicitly written the gauge fields, but they are easily computed. On the $d_2^2=d_1$ boundary the $A_t$ diverges and this might cause some consternation. But this divergence is actually a gauge artifact: the malady arises from a divergent constant and one can explicitly check that the field strengths remain finite at both boundaries.

%%%%%%%%%%%%%%%%%%%%%%%%%%%%%%%%%%%%%%%%%%%%%%%%%%%%%%%%%%%%%%%%%%%%%%%%%%%%%%%%%%%%%%%%%%%%%%%
\section*{Acknowledgments}

We thank Avik Chakraborty and Amitabh Virmani for discussions, and Sudhir Vempati for teaching us a simple trick that made it possible for Mathematica to do its job without choking.

%%%%%%%%%%%%%%%%%%%%%%%%%%%%%%%%%%%%%%%%%%%%%%%%%%%%%%%%%%%%%%%%%%%%%%%%%%%%%%%%
%%%%%%%%%%%%%%%%%%%%%%%%%%%%%%%%%%%%%%%%%%%%%%%%%%%%%%%%%%%%%%%%%%%%%%%%%%%%%%%%

\appendix

\section{The Boundary of the Basin}
\label{basindet}

Determining the boundary of the basin is in principle straightforward: one just has to demand the functions that define the fields should be regular and real for $0 < r <\infty$. But the problem splits up into many cases, so it is worthwhile being systematic. We will determine the basin here for the case of the rotating solution we constructed in section 3, we can reproduce the results we stated in the static case of section 2 by demanding that $\beta$, the rotation parameter, is zero. 

The first thing to note is that regularity is determined by the functions $f_1$ and $f_2$ defined in section 3. Regularity is equivalent to the statement that these functions should only have zeros away from the positive real axis. To avoid clutter, we will temperorily introduce the functions:  
\bea
f_1=a_1\ r^2+b_1\ r+1, \\
f_2=a_2\ r^2+b_2\ r+1,
\eea
and phrase our discussions in terms of them. At the end of the discussion\footnote{We will occasionally suppress the index $i$ in what follows, when we are making general statements valid for either $i$.} we will relate $a_{i=1,2}$ and $b_{i=1,2}$ to $d_1$ and $d_2$. These relations can be taken to be
\bea
a_1=\frac{2(d_2^2-d_1)}{1-\alpha}, \  \ b_1= -\frac{2 d_2}{1-\alpha}, \label{rel1} \\
a_2=\frac{2d_1}{1-\alpha}, \  \ b_2=-\frac{2 d_2}{1-\alpha}. \hspace{0.15in} \label{rel2}
\eea
where we have intrdoduced 
\bea
\alpha=\beta \cos \theta \label{alpha}
\eea
for convenience. Since $\beta =a_K/M_k <1$ and $|\cos \theta|<1$ we have the condition that $-1<\alpha<1$. 
 
Before we proceed, we make one comment: the end result for the attraction basin is simple and elegant, but embarassingly, we have not been able to derive it using a simple argument\footnote{Strictly speaking, our case-by-case argument gets complicated only when there is rotation. When there is no rotation, many of the separate cases collapse onto each other. Interestingly, the final result for the basin in terms of the $d_2 - d_1$ parameters is the {\em same} for both static and rotating black holes (\ref{statbasin}-\ref{statbound}). In particular it does not depend on the polar angle $\theta$ or the rotation parameter $\beta$.}. It seems evident that our result should follow from high-school notions about polynomials, roots, positivity, etc. but in the following, we have found it quicker to brute-force our way to the answer case-by-case.

First we notice that both $a_1$ and $a_2$ have to have the same sign. If not, for some large enough value of $r$, both $f_1/f_2$ and $f_1 f_2$ will become negative and run into trouble with the square roots in the solution. 

Now, if $a < 0$, the roots of $f$ cannot be complex: because $D= b^2-4a$ will be forced to be non-negative. On the other hand, when roots are real, we want them both to be negative so that there are no zeros on the positive real axis. For $a < 0$, this means that both
\bea
-b-\sqrt{b^2-4a} > 0, \ \ {\rm and} \ \ -b+\sqrt{b^2-4a}  > 0.
\eea
For negative $a$, it is clear that these equations cannot both be be true at the same time. This means that we must have $a$ positive. 

Now we discuss the case $a_i >0$ in detail. There are two ways in which the function $f$ can be regular on the entire positive real axis: it can either have compex (and mutually conjugate) roots, or both its roots are on the negative real line. Complex conjugate roots happen when $D \equiv b^2-4a <0$.  On the other hand, if the equations have real roots (ie., $D>0$), the bigger one will be
\bea
\frac{1}{2a}(-b_i+\sqrt{D_i}).
\eea 
We want this root to be $<0$. This forces $b > 0$. 

So in effect, what we have to do in order to determine the domain of attraction (the domain where the solutions are regular) is to consider the various possible combinations arising from $f_1$ and $f_2$. 
%between the roots of $f_1$ and $f_2$ where they both satisfy these conditions simultaneously. 
In principle there are four such cases ($a_i >0$ has to be also imposed for all of these cases as we already dicussed):
\begin{itemize}
\item $b_1^2-4a_1<0$ with $b_2^2-4a_2 <0$,
\item $b_1^2-4a_1<0$ with $b_2^2-4a_2 >0, b_2 >0$,
\item $b_1^2-4a_1>0, b_1>0$ with $b_2^2-4a_2 <0$,
\item $b_1^2-4a_1>0, b_1>0$ with $b_2^2-4a_2 >0, b_2 >0$.
\end{itemize}
We will call these the four branches of the basin. We will take the intersection of the regions in each bullet, and then take the union of the four regions to get the full attraction basin. 

In terms of (\ref{rel1}-\ref{rel2}), these four branches take the form 
\begin{itemize}
\item Branch 1:
\bea
d_2^2(2\alpha-1)+2 d_1 (1-\alpha)<0, \ \ d_2^2-d_1 >0, \\
d_2^2-2d_1(1+\alpha)<0, \ \ d_1 >0 \hspace{0.5in}
\eea
\item Branch 2:
\bea
d_2^2(2\alpha-1)+2 d_1 (1-\alpha)<0, \ \ d_2^2-d_1 >0, \\
d_2^2-2d_1(1+\alpha)>0,\ \ d_2<0, \ \ d_1 >0,  \hspace{0.2in}
\eea
\item Branch 3:
\bea
d_2^2(2\alpha-1)+2 d_1 (1-\alpha)>0, \ d_2<0, \ d_2^2-d_1 >0, \\
d_2^2-2d_1(1+\alpha)< 0, \ \ d_1 >0, \hspace{0.65in}
\eea
\item Branch 4:
\bea
d_2^2(2\alpha-1)+2 d_1 (1-\alpha)>0, \ d_2<0, \ d_2^2-d_1 >0, \\
d_2^2-2d_1(1+\alpha)> 0, \ \ d_2<0, \ \ d_1 >0, \hspace{0.45in}
\eea
\end{itemize}
At this stage we have to distinguish four cases again, because the qualitative features of the various branches depend on the ranges of $\alpha$! These are 
\bea
{\rm Case \ 1:}\  0<\alpha<\frac{1}{2}, \ \ {\rm Case\  2:}\
\frac{1}{2}<\alpha<1, \hspace{0.3in}\\
{\rm Case \ 3:}\ -\frac{1}{2}<\alpha<0, \ \
{\rm Case \ 4:}\ -1 <\alpha <-\frac{1}{2}. 
\eea
For each of these cases, one can investigate the four branches of the solution. The rest of the computation is merely an exercise in persistence, so we will merely present one illustrative subcase as an example and then quote the final result. The specific case we present is Branch 4 of $0 <\gamma \equiv -\alpha<\frac{1}{2}$. The contribution to the basin from that branch, arising from the intersection of the various inequalities, is presented in figure 3.
\begin{figure}
\begin{center}
\includegraphics[height=0.30\textheight]{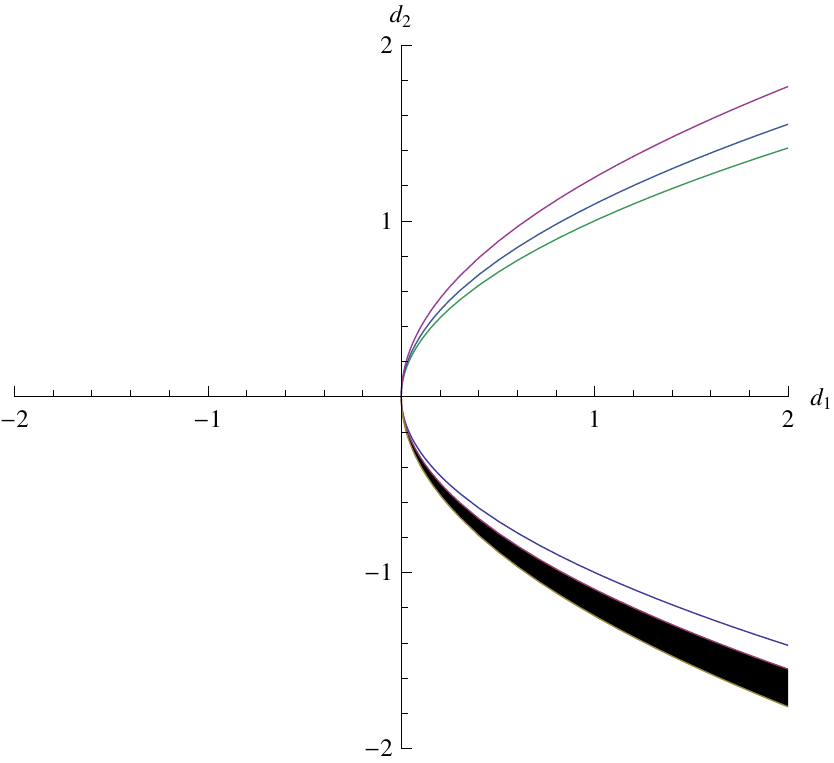}
\caption{The contribution to the basin from branch 4, for case 3. The curves that are plotted are the boundaries of the inequalities in the branch, and their intersection is the blackened region. More concretely, the branch 4 inequalities in terms of $\gamma =-\alpha$ take the form 
$ d_2^2<2 d_1 \Big(\frac{1+\gamma}{1+2 \gamma}\Big), d_2^2 > 2d_1(1-\gamma), d_2 <0, d_2^2> d_1, d_1 >0.$
We have made the plot above for the specific value $\gamma=0.4$. The ordering of the curves is fixed by the fact that $0<\gamma<\frac{1}{2}$.}
\end{center}
\end{figure}
Note that in the figure, the ordering of the curves depends crucially on the value of $\gamma$ and this is the reason why we had to work with the four  cases separately for the ranges of $\alpha$.

Once the dust settles, in each four of %the four cases for 
the ranges of $\alpha$, despite the vastly different intermediate steps, the union of all the four branches results in the {\em same} attraction basin:
\bea
d_2^2-d_1 >0, \ \ d_2<0, \ \ d_1 >0. 
\eea
In particular, this result is $\alpha$-independent for all values of $|\alpha|<1$. The boundaries of attraction are obtained when one replaces the first two inequalities with equalities, and retains the last one as an inequality. 

In principle, we need not have computed the basin separetely for negative $\alpha$, once we compute the basin for the range $0 < \alpha <1$. This is because of the symmetry presented in (\ref{sym1}-\ref{sym2}). %, which can be used to determine the basin by a symmetry argument. 
However, the intermediate results in the computation for positive and negative $\alpha$ are (at least superficially) quite different, so this works as a check of our results.

\section{One More Static Example}
\label{case3}

In this section, we will explore a case where the equations of motion are of the form (\ref{att1}-\ref{minansatze2}) but with the effective potential given by
\bea
V_{\rm eff}=e^{4 \phi}\ Q_1^2+e^{-6 \phi}\ Q_2^2.
\eea
Together with the resutls in \cite{A2} and the main body of this paper, this shows that the attraction basin can change completely depending on the system we consider.

The above specific form of the effective potential is another example where the resulting EOMs are integrable. This effective potential arises, for example, in an Einstein-Maxwell-dilaton system with one scalar and two vector fields (with magnetic charges $Q_1$ and $Q_2$ turned on for the gauge fields), with a gauge coupling matrix (see \cite{Goldstein, A1, A2}) given by
\bea
f_{ab}(\phi)=\left(
\begin{array}{cc}
e^{4 \phi}& 0 \\
0& e^{-6 \phi}
\end{array}
\right).
\eea
By relating to a Toda system, the coupled ODEs can be solved exactly. In the extremal limit, when the asymptotic value $a_0$ of $a(r)$ is set to unity, this was done in \cite{Goldstein}. We can easily redo their computations when $a_0$ is left arbitrary and express the final solutions in terms of two integration constants $d_1$ and $d_2$ similar to the Kaluza-Klein system in the main body of the paper. The result can be expressed in terms of two functions
\bea
A(r)&=&12\ (4 \ d_2^4 - 2 d_1\ d_2)\ r^4 - 12\ (4 d_2^3 - d_1)\ r^3 + 24 \ d_2^2 \ r^2 - 
  8 d_2\ r + 1, \\
B(r)&=&6\ d_1\ r^3 - 12\ d_2^2 \ r^2 + 6\ d_2\ r - 1.
\eea
In terms of these functions, the solution takes the form
\bea
\exp [10 \phi(r)] = \frac{3 Q_2^2}{2 Q_1^2} \frac{A(r)}{B(r)}, \ \ a(r)=\frac{2^{2/5} 6^{3/10}}{10^{1/2} Q_1^{3/5} Q_2^{2/5}}\frac{r}{A(r)^{1/5}B(r)^{1/10}}, \ \ b(r) =\frac{r}{a(r)}.
\eea

Demanding regularity in $0<r<\infty$ results in somewhat more complicated equations in this case, because the polynomials $A(r)$ and $B(r)$ are quartic and cubic and are best handled with a computer. But since there is no rotation, this is still very much a tractable situation. We will not present the details, but going about the regularity conditions systematically as in Appendix A results in a simple result 
\bea
2d_2^3 \ge d_1, \  \ d_1 \le 0, \ \ d_2 \le 0
\eea 
for the attraction basin. The boundaries of attraction are given by
\bea
2d_2^3 = d_1, \  \ d_1 = 0, \ {\rm with} \ d_2 < 0
\eea 
The basin structure is presented in figure 4.
\begin{figure}
\begin{center}
\includegraphics[height=0.40\textheight]{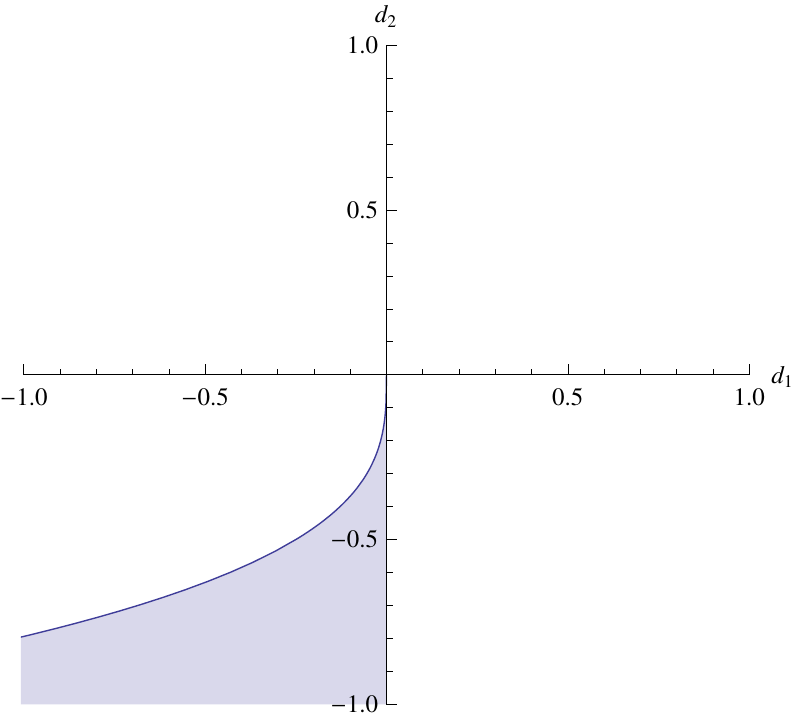}
\caption{Basin structure of this system should be contrasted to that of the KK black hole (figure 1) and those in \cite{A2}.}
\end{center}
\end{figure}

It is possible to do a near horizon perturbation theory analogous to that done in Appendix D of \cite{A2} for this system as well (cf. also section 2 of the present paper), to figure out the trajectory of a perturbed solution along the attraction basin. The result is that the basin gets sliced along
\bea
d_2= d_2^0  \label{slice3}
\eea
as was also the case in the solution we discussed in section 2. Here $d_2^0$ is a negative constant that corresponds to the $d_2$ of the unperturbed black hole. As in the case discussed in the main body of the paper, the leading scalar perturbation (call it $\phi_3$) is not at ${\cal O}(r)$, here it is at ${\cal O}(r^3)$. Perturbation theory tells us the relations
\bea
d_1=d_1^0+\frac{1}{6} \phi_3, \ \ d_2=d_2^0,
\eea
upon matching the perturbed solution to the parameters of the new solution. Eliminating $\phi_3$ leads to (\ref{slice3}). 

It is trivial to check that the warp factors defined via $\Delta = r^4/a(r)^4$ goes at the $d_1=0$ boundary as $r^{8/5}$ and at the $2d_2^3=d_1$ as $r^{18/5}$. This geometry is another version of the extremal subtracted (subttractor) geometry.

For the specific case where $d_2^0=-2$, we present the curves that arise for a selection of values of $d_1$, as we dial it from one side of the boundary 1, through boundary 1, through the basin, through boudnary 2, to the other side of boundary 2.  
\begin{figure}
\begin{center}
\includegraphics[height=0.30\textheight]{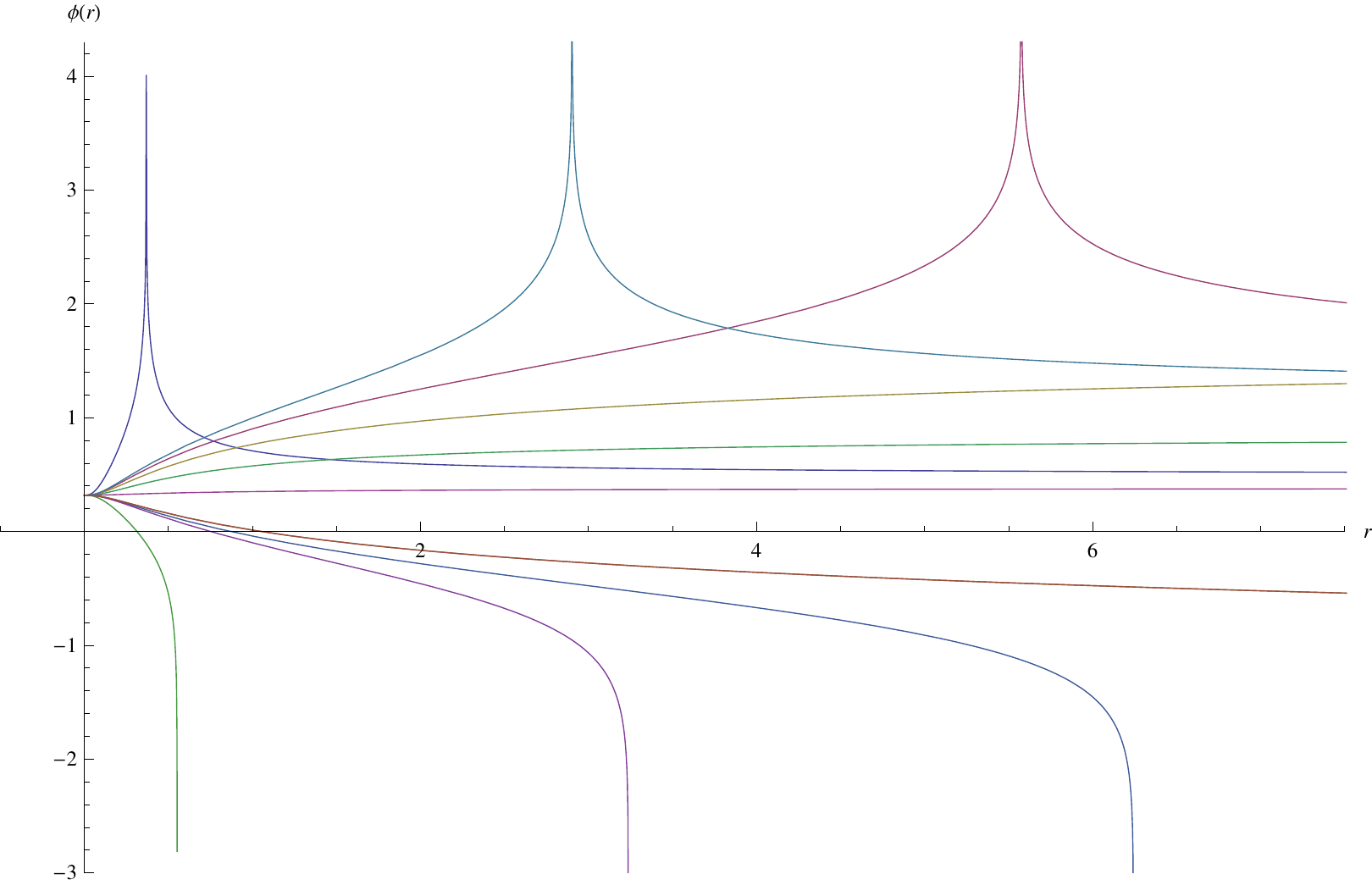}
\caption{The radial profiles of the scalars in the model in Appendix 2. The strange shapes of the curves that start beyond the upper boundary of the basin are perfectly legitimate features of the plot.}
\end{center}
\end{figure}
The nature of the divergent solutions is somewhat different here\footnote{More precisely, the curves beyond the upper boundary have divergences, but they asymptote to constant scalalr values at inifinity.} than in the other cases we have examined. But this is nothing pathological. The new feature is that the asymptotic values of the fields in this solution do not uniquely fix an ordered pair of {\em real} numbers $(d_1, d_2)$ in this case. In the previous cases, the only real pair $(d_1, d_2)$ that gave rise to a given set of asymptotic data, was the non-divergent one. Here on the other hand, while one of the solutions falls inside the basin and gives rise to the expected curve, another one lies outside and is seen as a divergence before it crosses inside the basin. Of course this is not a regular solution, and as expected, in the near-horizon region which is responsible for the attraction mechanism, it lies outside the basin. 

%The boundary of the attraction basin is at $2d_2^3 = d_1$ and $d_1 = 0$. It is trivial to check that the warp factors defined via $\Delta = r^4/a(r)^4$ goes at both boundaries as $r^{12/5}$. This is another version of the extremal subtracted (subttractor) geometry.

%%%%%%%%%%%%%%%%%%%%%%%%%%%%%%%%%%%%%%%%%%%%%%%%%%%%%%%%%%%%%%%%%%%%%%%%%%%%%%%%%%%%%%%%%%%%%%%%%%%%%%

% ==========================================================================
%
%%%%%%%%%%%%%%%%%%%%%%%%%%%%%%%%%%%%%%%%%%%%%%%%%%%%%%%%%%%%%%%%%%%%%%%%%%%%
%                      REFERENCES                            %
%%%%%%%%%%%%%%%%%%%%%%%%%%%%%%%%%%%%%%%%%%%%%%%%%%%%%%%%%%%%%%%%%%%%%%%%%%%%
%\newpage
%\bibliography{metasusy}

\begin{thebibliography}{19}        %here 19 is the widest mark...
%-----Type it \bibitem[how it is marked]{how we call it}Authors,
%-----Citations are then made by \cite{how we call it} in text
%-----\bibitem without [how it is denoted] is numbered 1,2,3....

%\cite{Sen:1995in}
%\bibitem{SenSmall}  A.~Sen,  ``Extremal black holes and elementary string states,''  Mod.\ Phys.\ Lett.\ A {\bf 10}, 2081 (1995)  [hep-th/9504147].
  %%CITATION = HEP-TH/9504147;%%

%\bibitem{StromingerVafa}   A.~Strominger and C.~Vafa,``Microscopic origin of the Bekenstein-Hawking entropy,''Phys.\ Lett.\ B {\bf 379}, 99 (1996) [hep-th/9601029].
  %%CITATION = HEP-TH/9601029;%%


%\cite{Ferrara:1995ih}
\bibitem{FKS} 
  S.~Ferrara, R.~Kallosh and A.~Strominger,
  ``N=2 extremal black holes,''
  Phys.\ Rev.\ D {\bf 52}, 5412 (1995)
  [hep-th/9508072].
  %%CITATION = HEP-TH/9508072;%%

  A.~Sen,
  ``Black Hole Entropy Function, Attractors and Precision Counting of Microstates,''
  Gen.\ Rel.\ Grav.\  {\bf 40}, 2249 (2008)
  [arXiv:0708.1270 [hep-th]].
  %%CITATION = ARXIV:0708.1270;%%
  %\cite{Banerjee:2009uk}
  
\bibitem{A1}
  A.~Chakraborty and C.~Krishnan,
  ``Subttractors,''
  arXiv:1212.1875 [hep-th].
  %%CITATION = ARXIV:1212.1875;%%
%\cite{Sen:2007qy}

\bibitem{A2}
  A.~Chakraborty and C.~Krishnan,
  ``Attraction, with Boundaries,''
  arXiv:1212.6919 [hep-th].
  %%CITATION = ARXIV:1212.1875;%%
%\cite{Sen:2007qy}

\bibitem{CL1} 
  M.~Cvetic and F.~Larsen,
  ``Conformal Symmetry for General Black Holes,''
  JHEP {\bf 1202}, 122 (2012)
  [arXiv:1106.3341 [hep-th]].
  %%CITATION = ARXIV:1106.3341;%%


%\cite{Cvetic:2011dn}
\bibitem{CL2} 
  M.~Cvetic and F.~Larsen,
  ``Conformal Symmetry for Black Holes in Four Dimensions,''
  JHEP {\bf 1209}, 076 (2012)
  [arXiv:1112.4846 [hep-th]].

%\cite{Cvetic:2012tr}
\bibitem{CG} 
  M.~Cvetic and G.~W.~Gibbons,
  ``Conformal Symmetry of a Black Hole as a Scaling Limit: A Black Hole in an Asymptotically Conical Box,''
  JHEP {\bf 1207}, 014 (2012)
  [arXiv:1201.0601 [hep-th]].
  %%CITATION = ARXIV:1201.0601;%%

\bibitem{KerrCFT} 
  M.~Guica, T.~Hartman, W.~Song and A.~Strominger,
  ``The Kerr/CFT Correspondence,''
  Phys.\ Rev.\ D {\bf 80}, 124008 (2009)
  [arXiv:0809.4266 [hep-th]].
  %%CITATION = ARXIV:0809.4266;%%

H.~Lu, J.~Mei and C.~N.~Pope,
  ``Kerr/CFT Correspondence in Diverse Dimensions,''
  JHEP {\bf 0904}, 054 (2009)
  [arXiv:0811.2225 [hep-th]].
  %%CITATION = ARXIV:0811.2225;%%
  
T.~Hartman, K.~Murata, T.~Nishioka and A.~Strominger,
  ``CFT Duals for Extreme Black Holes,''
  JHEP {\bf 0904}, 019 (2009)
  [arXiv:0811.4393 [hep-th]].
  %%CITATION = ARXIV:0811.4393;%%

C.~Krishnan and S.~Kuperstein,
  ``A Comment on Kerr-CFT and Wald Entropy,''
  Phys.\ Lett.\ B {\bf 677}, 326 (2009)
  [arXiv:0903.2169 [hep-th]].
  %%CITATION = ARXIV:0903.2169;%%

%\cite{Castro:2010fd}
\bibitem{HCS} 
  A.~Castro, A.~Maloney and A.~Strominger,
  ``Hidden Conformal Symmetry of the Kerr Black Hole,''
  Phys.\ Rev.\ D {\bf 82}, 024008 (2010)
  [arXiv:1004.0996 [hep-th]].
  %%CITATION = ARXIV:1004.0996;%%

%\cite{Krishnan:2010pv}
%\bibitem{CK} 
  C.~Krishnan,
  ``Hidden Conformal Symmetries of Five-Dimensional Black Holes,''
  JHEP {\bf 1007}, 039 (2010)
  [arXiv:1004.3537 [hep-th]].
  %%CITATION = ARXIV:1004.3537;%%
  
  G.~Compere,
  ``The Kerr/CFT correspondence and its extensions: a comprehensive review,''
  Living Rev.\ Rel.\  {\bf 15}, 11 (2012)
  [arXiv:1203.3561 [hep-th]].
  %%CITATION = ARXIV:1203.3561;%%

%\cite{Yuan:2013ts}
\bibitem{General1}
  F.~-F.~Yuan and Y.~-C.~Huang,
  ``Harrison metrics for the Schwarzschild black hole,''
  arXiv:1301.6548 [hep-th].
  %%CITATION = ARXIV:1301.6548;%%

%\cite{Keeler:2012mq}
%\bibitem{General2} 
  C.~Keeler and F.~Larsen,
  ``Separability of Black Holes in String Theory,''
  JHEP {\bf 1210}, 152 (2012)
  [arXiv:1207.5928 [hep-th]].
  %%CITATION = ARXIV:1207.5928;%%


\bibitem{Pope} C. Pope, ``Kaluza-Klein Theory," http://people.physics.tamu.edu/pope/ihplec.pdf


%\cite{Goldstein:2005hq}
\bibitem{Goldstein} 
  K.~Goldstein, N.~Iizuka, R.~P.~Jena and S.~P.~Trivedi,
  ``Non-supersymmetric attractors,''
  Phys.\ Rev.\ D {\bf 72}, 124021 (2005)
  [hep-th/0507096].
  %%CITATION = HEP-TH/0507096;%%


%\cite{Rasheed:1995zv}
\bibitem{Rasheed} 
  D.~Rasheed,
  ``The Rotating dyonic black holes of Kaluza-Klein theory,''
  Nucl.\ Phys.\ B {\bf 454}, 379 (1995)
  [hep-th/9505038].
  %%CITATION = HEP-TH/9505038;%%
  %108 citations counted in INSPIRE as of 12 Mar 2013
%\cite{Cvetic:1997uw}

%\cite{Matos:1996km}
\bibitem{Matos} 
  T.~Matos and C.~Mora,
  ``Stationary dilatons with arbitrary electromagnetic field,''
  Class.\ Quant.\ Grav.\  {\bf 14}, 2331 (1997)
  [hep-th/9610013].
  %%CITATION = HEP-TH/9610013;%%
  %19 citations counted in INSPIRE as of 12 Mar 2013

\bibitem{LarsenKK} 
  F.~Larsen,
  ``Rotating Kaluza-Klein black holes,''
  Nucl.\ Phys.\ B {\bf 575}, 211 (2000)
  [hep-th/9909102].
  %%CITATION = HEP-TH/9909102;%%
  %71 citations counted in INSPIRE as of 12 Mar 2013

 %\cite{Astefanesei:2006dd}
\bibitem{Kevin} 
  D.~Astefanesei, K.~Goldstein, R.~P.~Jena, A.~Sen and S.~P.~Trivedi,
  ``Rotating attractors,''
  JHEP {\bf 0610}, 058 (2006)
  [hep-th/0606244].
  %%CITATION = HEP-TH/0606244;%%


%\cite{Cvetic:2013cja}
\bibitem{CveticStudent} 
  M.~Cvetic, M.~Guica and Z.~H.~Saleem,
  ``General black holes, untwisted,''
  arXiv:1302.7032 [hep-th].
  %%CITATION = ARXIV:1302.7032;%%
  %1 citations counted in INSPIRE as of 12 Mar 201


%\cite{Baggio:2012db}
\bibitem{JdBoer} 
  M.~Baggio, J.~de Boer, J.~I.~Jottar and D.~R.~Mayerson,
  ``Conformal Symmetry for Black Holes in Four Dimensions and Irrelevant Deformations,''
  arXiv:1210.7695 [hep-th].
  %%CITATION = ARXIV:1210.7695;%%

%\cite{Andrianopoli:2013kya}
\bibitem{Torino} 
  L.~Andrianopoli, R.~D'Auria, A.~Gallerati and M.~Trigiante,
  ``Extremal Limits of Rotating Black Holes,''
  arXiv:1303.1756 [hep-th].
  %%CITATION = ARXIV:1303.1756;%%


\bibitem{Amitabh} 
  A.~Virmani,
  ``Subtracted Geometry From Harrison Transformations,''
  JHEP {\bf 1207}, 086 (2012)
  [arXiv:1203.5088 [hep-th]].
  %%CITATION = ARXIV:1203.5088;%%

  %\cite{Bertini:2011ga}
\bibitem{Klemm} 
  S.~Bertini, S.~L.~Cacciatori and D.~Klemm,
  ``Conformal structure of the Schwarzschild black hole,''
  Phys.\ Rev.\ D {\bf 85}, 064018 (2012)
  [arXiv:1106.0999 [hep-th]].
  %%CITATION = ARXIV:1106.0999;%%


\bibitem{General} 
  M.~Cvetic and D.~Youm,
  ``General rotating five-dimensional black holes of toroidally compactified heterotic string,''
  Nucl.\ Phys.\ B {\bf 476}, 118 (1996)
  [hep-th/9603100].
  %%CITATION = HEP-TH/9603100;%%
 %\cite{Cvetic:1995kv}

  M.~Cvetic and D.~Youm,
  ``All the static spherically symmetric black holes of heterotic string on a six torus,''
  Nucl.\ Phys.\ B {\bf 472}, 249 (1996)
  [hep-th/9512127].
  %%CITATION = HEP-TH/9512127;%%

  M.~Cvetic and F.~Larsen,
  ``General rotating black holes in string theory: Grey body factors and event horizons,''
  Phys.\ Rev.\ D {\bf 56}, 4994 (1997)
  [hep-th/9705192].
  %%CITATION = HEP-TH/9705192;%%  

\bibitem{GibMaed} 
  G.~W.~Gibbons and K.~-i.~Maeda,
  ``Black Holes and Membranes in Higher Dimensional Theories with Dilaton Fields,''
  Nucl.\ Phys.\ B {\bf 298}, 741 (1988).
  %%CITATION = NUPHA,B298,741;%%

  %\cite{Lu:1996jr}
\bibitem{Pope2} 
  H.~Lu and C.~N.~Pope,
  ``SL(N+1,R) Toda solitons in supergravities,''
  Int.\ J.\ Mod.\ Phys.\ A {\bf 12}, 2061 (1997)
  [hep-th/9607027].
  %%CITATION = HEP-TH/9607027;%%
  
%\bibitem{K}``Generalized Subtracted Geometries and the Phases of Attraction", to appear.

%\cite{Arean:2010zw}
\bibitem{SCK1} 
  D.~Arean, P.~Basu and C.~Krishnan,
  ``The Many Phases of Holographic Superfluids,''
  JHEP {\bf 1010}, 006 (2010)
  [arXiv:1006.5165 [hep-th]].
  %%CITATION = ARXIV:1006.5165;%%
  %24 citations counted in INSPIRE as of 12 Mar 2013

%\cite{Arean:2010wu}
\bibitem{SCK2} 
  D.~Arean, M.~Bertolini, C.~Krishnan and T.~Prochazka,
  ``Type IIB Holographic Superfluid Flows,''
  JHEP {\bf 1103}, 008 (2011)
  [arXiv:1010.5777 [hep-th]].
  %%CITATION = ARXIV:1010.5777;%%
  %11 citations counted in INSPIRE as of 12 Mar 2013

%\cite{Arean:2011gz}
\bibitem{SCK3} 
  D.~Arean, M.~Bertolini, C.~Krishnan and T.~Prochazka,
  ``Quantum Critical Superfluid Flows and Anisotropic Domain Walls,''
  JHEP {\bf 1109}, 131 (2011)
  [arXiv:1106.1053 [hep-th]].
  %%CITATION = ARXIV:1106.1053;%%
  %5 citations counted in INSPIRE as of 12 Mar 2013

%\cite{Krishnan:2010un}
%\bibitem{CKreview}  C.~Krishnan,``Quantum Field Theory, Black Holes and Holography,''arXiv:1011.5875 [hep-th].
  %%CITATION = ARXIV:1011.5875;%%
%\cite{Gibbons:1987ps}


  
\end{thebibliography}

\end{document}